\documentclass[fleqn,usenatbib]{mnras}

\usepackage{newtxtext,newtxmath}

\usepackage[T1]{fontenc}
\usepackage{orcidlink}

\DeclareRobustCommand{\VAN}[3]{#2}
\let\VANthebibliography\thebibliography
\def\thebibliography{\DeclareRobustCommand{\VAN}[3]{##3}\VANthebibliography}

\usepackage{graphicx}	
\usepackage{amsmath}	

\newcommand{\msun}{{\,\rm M_\odot}}

\newcommand{\kms}{\,{\rm km}\,{\rm s}^{-1}}

\newcommand{\kpc}{\,{\rm kpc}}

\def\gsim{ \lower .75ex \hbox{$\sim$} \llap{\raise .27ex \hbox{$>$}} }
\def\lsim{ \lower .75ex \hbox{$\sim$} \llap{\raise .27ex \hbox{$<$}} }

\title[Satellite disc disruption]{The impact of disc disruption on Milky Way satellite counts}

\author[M. R. Lovell et al.]{
Mark R. Lovell$^{\orcidlink{0000-0001-5609-514X}
}$,$^{1,2}$\thanks{E-mail: m.r.lovell@durham.ac.uk}
Alexander H.~Riley$^{\orcidlink{0000-0001-5805-5766}}$,$^{1,2,3}$ and
Isabel M.E. Santos-Santos$^{\orcidlink{0000-0001-6054-2897}}$$^{1,2}$
\\
$^{1}$ Institute for Computational Cosmology, Durham University, South Road, Durham DH1 3LE, United Kingdom\\
$^{2}$ Department of Physics, Durham University, South Road, Durham DH1 3LE, United Kingdom\\
$^{3}$ Lund Observatory, Division of Astrophysics, Department of Physics, Lund University, SE-221 00 Lund, Sweden}

\date{Accepted XXX. Received YYY; in original form ZZZ}

\pubyear{2026}

\begin{document}
\label{firstpage}
\pagerange{\pageref{firstpage}--\pageref{lastpage}}
\maketitle

\begin{abstract}
Estimates for the total number of Milky Way (MW) satellites are often generated from a combination of the observed number of satellites in surveys, adjustments for the completeness of those surveys, and theoretical expectations from halo assembly modelling. One of the features of this modelling is disruption by the MW stellar disc. We examine the effect of degrees of disc disruption on inferred satellite counts, by means of an N-body simulation of a MW-mass halo plus a toy model for this disruption. We use a fictional all-sky survey to show that high resilience to disc disruption predicts small populations of satellites that are radially very concentrated around the central galaxy and are hosted by massive subhaloes, while low resilience predicts many more satellites with a less concentrated radial distribution and hosted within less massive subhaloes. We show that the most massive subhaloes are particularly susceptible to disruption due to their radial orbits, and in their putative absence galaxy formation must occur in lower mass haloes that have a shallower radial number density profile.  We then demonstrate this phenomenon for a combination of the Pan-STARRS and DES surveys. It is therefore necessary to account for uncertainty in the disc disruption radius when making predictions for MW satellite distributions.     
\end{abstract}

\begin{keywords}
Local Group -- galaxies:dwarf -- dark matter
\end{keywords}



\section{Introduction}

The Milky Way (MW) satellite galaxy system provides a key set of observables with which to constrain cosmology and astrophysics models, whether in the application of astrophysical processes -- supernova feedback, reionization heating \citep{Bullock_00,Benson03} -- or in the impact of novel dark matter physics. These observables include the number of satellites \citep{Klypin99,Moore99,Polisensky11,Lovell14,Kennedy14,Cherry17,Newton18,Nadler21}, their halo mass--stellar mass relation \citep{Walker09,BoylanKolchin11,BoylanKolchin12}, their spatial distribution \citep{Newton18} and their stellar ages \citep{Maccio19,Lovell20c}. Any comprehensive model of satellite galaxy formation much match all of these observations simultaneously in order to be considered viable.

The MW satellite counts are arguably the cleanest test of dark matter models, in that the number of dark matter subhaloes is in principle readily obtainable from simulations of sufficient spatial resolution, and can then be compared to observations at various resolution levels. As a first step, one can compare model predictions of subhalo counts to the number of detected MW satellites for extremely conservative constraints on the model parameters. The second, follow-up procedure is to extrapolate the total number of MW satellites from an understanding  of current surveys' sky coverage and depth along with the expectations for dark matter subhalo masses and spatial distributions; for our purposes we define `total' estimate as the number of satellites within 300~kpc of the centre of the MW, and label this quantity $n_\rmn{est}$ throughout this paper. A series of astrophysical and statistical models have been used to infer the number of satellites, and they disagree by a factor of 2 \citep{Newton18,Kim18,Nadler21,Manwadkar:2022,Weerasooriya:2023}, thus the procedure for performing this extrapolation is highly uncertain.

One of the key uncertainties affecting these estimates is the rate of subhalo disruption, first by the MW dark matter halo and second by the stellar disc. A subhalo whose dynamical mass at infall is sufficiently close to that of the MW will sink to the centre and merge with the latter within a Hubble time, thus the satellite it hosts will likewise merge onto the MW stellar halo and not contribute towards the satellite count \citep[e.g.][]{Lacey93,Simha17}. Smaller satellites are known to disappear in N-body simulations for numerical reasons rather than due to physical processes, either because the subhalo finder is unable to identify the density peak as self-bound when close to the host halo centre \citep{Onions12,ForouharMoreno25}, or because the finite simulation mass resolution can lead to premature mass loss, and even spurious disruption pericentre \citep{JiangF15,Lovell25}. Some studies have argued that with infinite resolution it should be impossible for a host halo to disrupt a subhalo in its entirety, and so a heavily stripped yet self-bound remnant should orbit around the host galaxy in perpetuity \citep{Errani23}.

While the MW's dark matter halo alone may well not be able to destroy satellites below a given subhalo-to-host mass ratio, the baryonic disc of stars and gas coult potentially to do so given its higher mass concentration and how it induces adiabatic contraction of the host halo \citep{Blumenthal86,Gnedin04}. Objects such as the Gaia Sausage/Enceladus are considered to be the remnant of a MW satellite disrupted in the distant past \citep{Belokurov18,Helmi18}, the Sagittarius dwarf spheroidal is undergoing a similar process at the present day \citep{Velazquez95}, and there is evidence that the low surface brightness structures of the Crater~II and Antlia~II satellites were induced by host interactions \citep{Sanders:2018, JiA21, Limberg:2025, Vivas:2026, Atzberger:2026}. The physically disrupted stellar material will subsequently contribute to the observed exsitu stellar halo. The question then follows of how to obtain the correct disruption rate -- negating numerical disruption from the halo while modelling physical disruption from the disc accurately -- with simulations of limited resolution \citep[ee e.g.][]{Grand21}, and what impact these modelling choices will have on the number of satellite galaxies we infer.

Studies that have considered this problem include \citet{Newton18} and \citet{Nadler20}, in both cases as part of broader attempts to estimate $n_\rmn{est}$. The latter employs machine learning techniques trained on a subset of the Latte hydrodynamical simulations \citep{GarrisonKimmel17} to ascertain which satellites are likely to be destroyed as per their orbital parameters \citep{Nadler18}. They then allow the strength of their disruption model to vary as part of a wider statistical model fitted to the Dark Energy Survey \citep[DES;][]{Bechtol15,DrlicaWagner15} and Pan-STARRS \citep{Laevens15} footprint satellite counts. \citet{Newton18} instead use the APOSTLE simulations \citep{Fattahi16,Sawala16a} to infer when satellites are disrupted. The net result of these two approaches -- including multiple processes and model features beyond disc disruption -- produces very different satellite counts, with $\sim220$ for \citet{Nadler21} and $\sim120$ for \citet{Newton18}, therefore there remains considerable uncertainty in the correct approach.

One can begin to assess from where the differences in the satellite count arise by considering other predictions for each model. One such property is the radial distribution of satellites about the MW centre, which is sensitive to uncertainty in the rate of disc disruption. If excessive disc disruption is present in the physical model, there is a risk that the model will spuriously destroy objects that are on closer orbits, that formed earlier, and were more massive at infall \citep{Lovell21}, all of which are properties that make a subhalo more likely to generate a bright, detectable galaxy. If one applies a statistical fit to the observed satellite counts -- but not to their radial distribution -- the fit will compensate for this disruption by placing galaxies in subhaloes that are more distant and thus likely to be of lower mass, which are much more numerous beyond the survey depth footprint and therefore the total inferred satellite count may be too high compared to an accurate disc disruption model. With respect to the inferred satellite number counts studies, the \citet{Newton18} model results give a reasonable agreement with current estimates of the satellite radial distribution, whereas \citet{Nadler20} find their distribution is less concentrated than the observed counterpart within the two survey footprints. \citet{Pham23} similarly found that specifically the \citet{Nadler18} disc disruption model produced a radial distribution of the brightest 13 satellites that is significantly less concentrated than that of the MW.

We investigate this problem by developing a very simple toy model of disc disruption and applying the result to an $N$-body simulation. We will simplify the numerous complexities of satellite surveys by conceiving of a fictional survey in which we can have absolute control of the systematics. We combine this survey and the $N$-body simulation to generate an intuition for how changing the degree of disc disruption will impact both the expected number of galaxies and the radial distribution of galaxies. We will further apply the DES and Pan-STARRS selection functions to the simulation to make for a more realistic comparison, while cautioning that our simulated halo is only MW-like in its mass and not in its environment, and therefore is a purely qualitative rather than quantitative comparison.

This paper is organised as follows. In Section~\ref{sec:sim} we present the simulation used in this paper, including the generation of its initial conditions, and in Section~\ref{sec:mod} we present the disc disruption model and the notional survey to which it is applied. Our results from the fictional survey are shown in Section~\ref{sec:50kpc}, and our DES/Pan-STARRS derived results are in Section~\ref{sec:despan}. We draw our conclusions in Section~\ref{sec:conc}. 

\begin{figure*}
    \centering
    \includegraphics[scale=0.23]{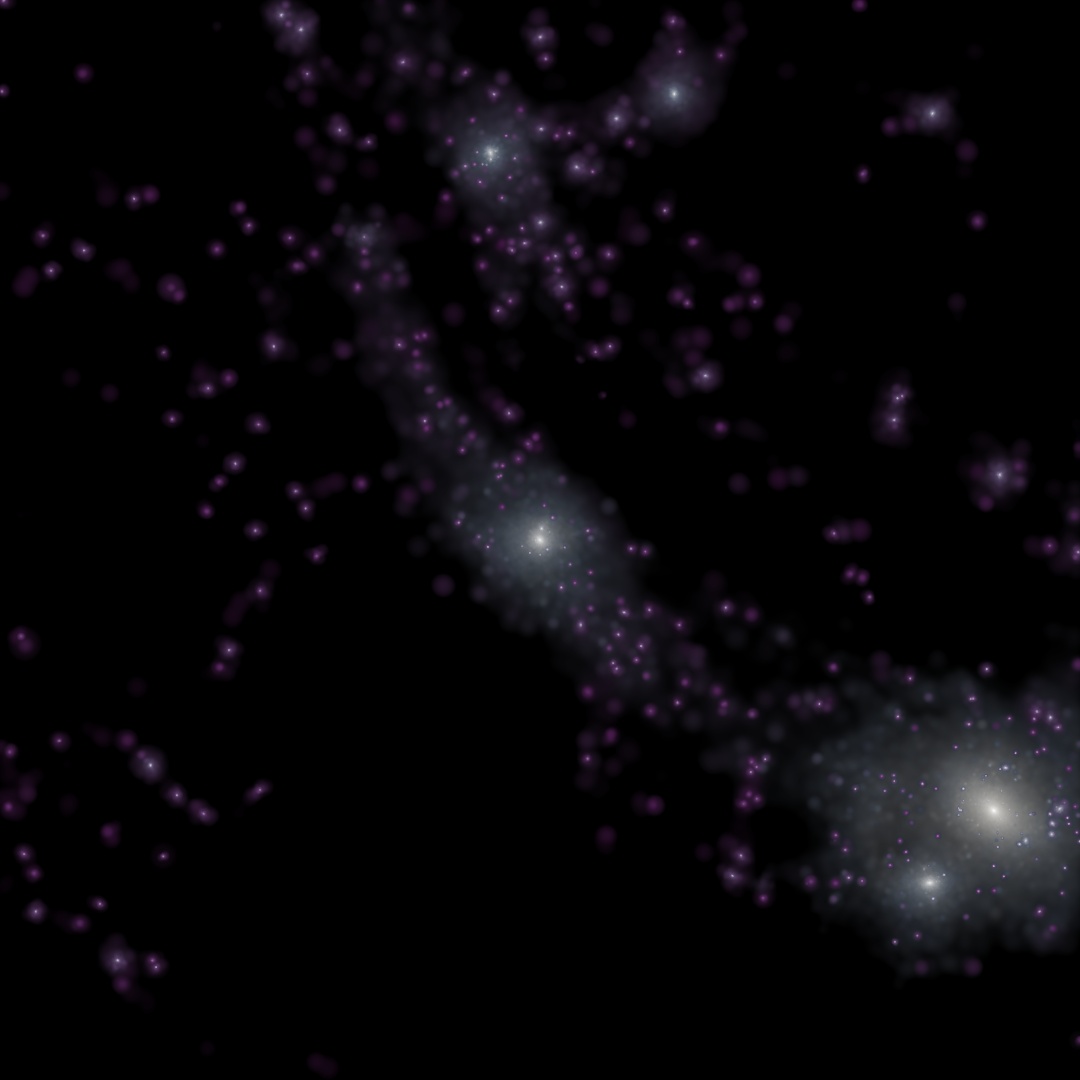}
     \includegraphics[scale=0.23]{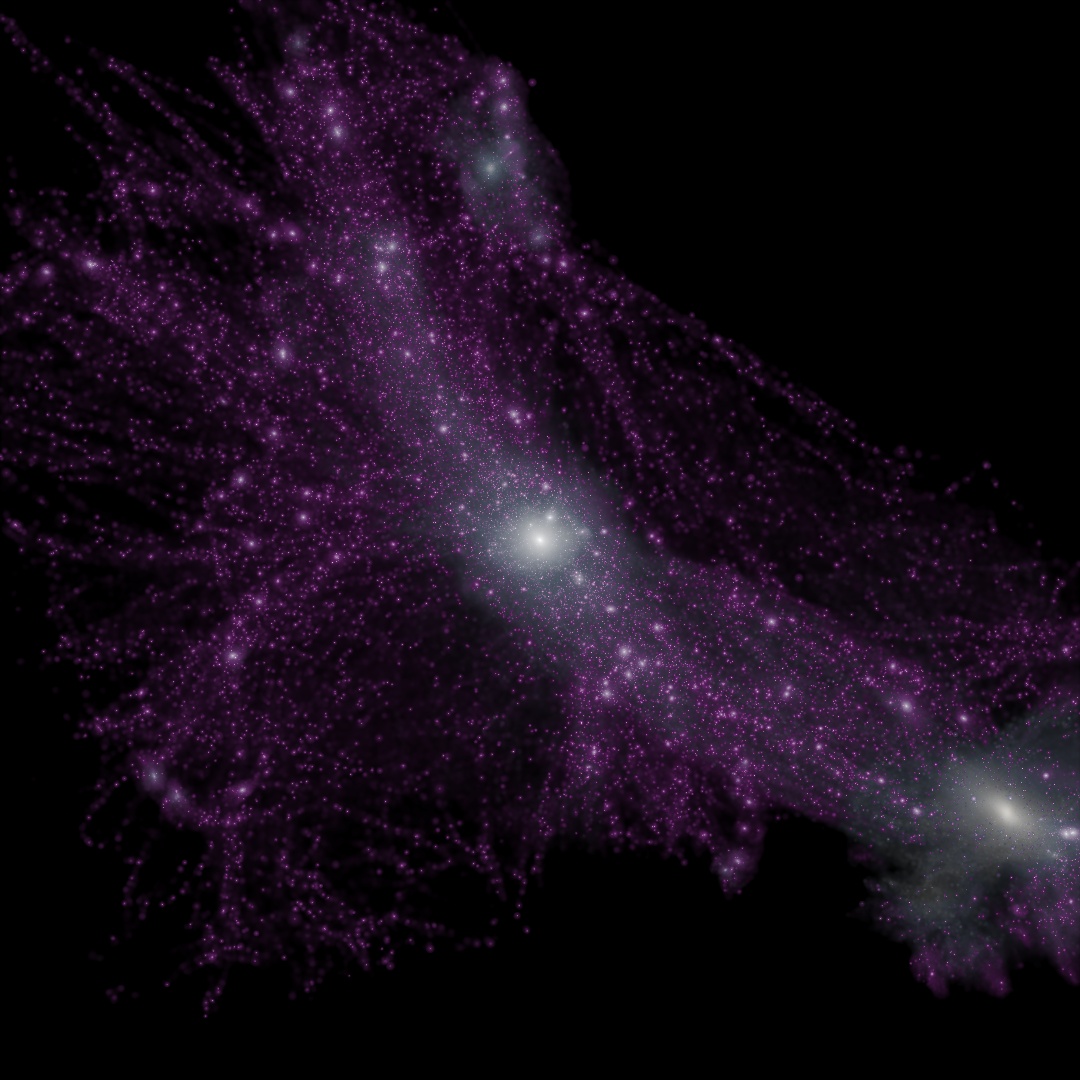}
    \caption{Images of the halo used in this study at $z=0$. In the left-hand panel we show the original halo within the EAGLE DMO-100Mpc box, and in the right-hand panel we present our zoomed resimulation. Each image slice is 4~$h^{-1}\rmn{Mpc}$ wide and 2~$h^{-1}\rmn{Mpc}$ thick. The image intensity indicates the density and hue the velocity dispersion, with purple for velocity dispersion $\lsim5$~$\kms$ and yellow for velocity dispersion $\gsim200$~$\kms$.}
    \label{fig:images}
\end{figure*}

\section{Simulations}
\label{sec:sim}

Our goal for this study is to follow the orbits of satellite galaxies around a MW-mass halo. We therefore identify a halo of the appropriate mass that is suitable for re-simulation with the zoom-in technique, generate its initial conditions at the required mass resolution, and perform the simulation with a high degree of time resolution. We demonstrate our process below.

An ideal MW-analogue, dark matter-only halo simulation would consist of a halo that has a dynamical mass within current bounds, $\sim[0.8,1.2]\times10^{12}$~$\msun$ \citep[e.g.][]{Callingham19}, and is located within a larger MW-like environment, with a Large Magellanic Cloud (LMC)-mass companion at a distance of $\sim50$~kpc from the MW centre, an M31-mass companion at $\sim$750~kpc distant, and a broader local environment that attempts to match the local large scale structure and matter density. This is a major task, as undertaken for the CLUES \citep{Libeskind10}, HESTIA \citep{Libeskind20}, and SIBELIUS \citep{Sawala22} projects, and is beyond the scope of this study, both for the level of detail required and for the computational expense and storage requirements of running a full Local Group (LG) simulation including a M31-analogue. We will therefore restrict ourselves to an isolated MW-mass halo, the parameters of which we describe below, and state that a comprehensive assessment of the MW satellite counts will require a cosmological environment more akin to the LG.    

We set our MW-mass halo requirements as:

\begin{itemize}
    \item Virial mass $M_{200}=[0.8,1.2]\times10^{12}$~$\msun$
    \item Formation redshift $z_\rmn{form}>1$
    \item Most massive companion within 3~Mpc dynamical mass $<5\times10^{11}$~$\msun$
    \item Total mass enclosed within 3~Mpc of the halo centre $<1\times10^{13}$~$\msun$
\end{itemize}

\noindent
which, to reiterate our discussion above, are selected to exclude expensive M31-analogue counterparts; we define $z_\rmn{form}$ as the redshift at which the primary progenitor mass attains half of its present day mass. We select suitable candidates using the CosmICweb online service \citep[https://cosmicweb.eu/;][]{Buehlmann24}, which writes parameter files for the MUSIC initial conditions code \citep{Hahn11} to generate zoomed initial conditions. CosmICweb supplies initial conditions for a range of simulations: we opt to use the EAGLE DMO-100Mpc box with particle mass $1.2\times10^{7}$~$\msun$ \citep{Schaye15}, for which three haloes meet the MW-analogue requirements listed above. We selected one of these three haloes for resimulation, labelled 29709583 in the CosmICWeb catalogue, and used CosmICweb to generate a MUSIC parameter file with a redshift $z=127$ and a maximum resolution level of 13, which creates a high-resolution region of particle mass $7.1\times10^{4}$~$\msun$, or $162\times$ better than the parent simulation. We adopt a softening length of 148~pc. The cosmological parameters are the same as for the parent simulation, and are consistent with the \citet{PlanckCP13} results: $\Omega_\rmn{m}=0.307$, $\Omega_{\Lambda}=0.693$, $H_{0}=67.77$~$\kms\rmn{Mpc}^{-1}$, $\sigma_{8}=0.8288$ and $n_\rmn{s}=0.9611$.

The initial conditions are evolved forward with the {\sc arepo} code \citep{Springel10,Weinberger20} through to the present day. We write 1024 snapshots to disk across time in order to follow the subhalo orbits with suitable time resolution. In Fig.~\ref{fig:images} we present an image of this halo at the present day, together with an image of the same halo from the parent box.


The large scale structure is identical between images, with an arc of material starting at the top-left and ending at the bottom right in a more massive halo that appears to be at the edge of the high resolution region. The apparent size and velocity dispersion of the central halo is replicated, and we have also verified that the halo mass functions within the mutually-resolved region are the same (not shown). We are therefore confident that our simulation is an accurate high resolution resimulation of the original EAGLE  DMO-100Mpc halo. 

Subhaloes are identified using the {\sc subfind} gravitational unbinding code \citep{Springel01}. All subhaloes that are accreted across the MW-progenitor 300~kpc (comoving) distance threshold are considered to be potential hosts of MW satellites. We identify each subhalo's most-bound particle at this accretion time and use this for the satellite present-day position; in this way we compensate for when such subhaloes may be either lost by the halo finder of spuriously destroyed due to limited resolution as discussed above. Note that we do not remove subhaloes that have likely merged onto the host under dynamical friction, these are instead removed implicitly by the disruption model as merging haloes will have very small final apocentres. 

\begin{figure*}
    \centering
    \includegraphics[scale=0.35]{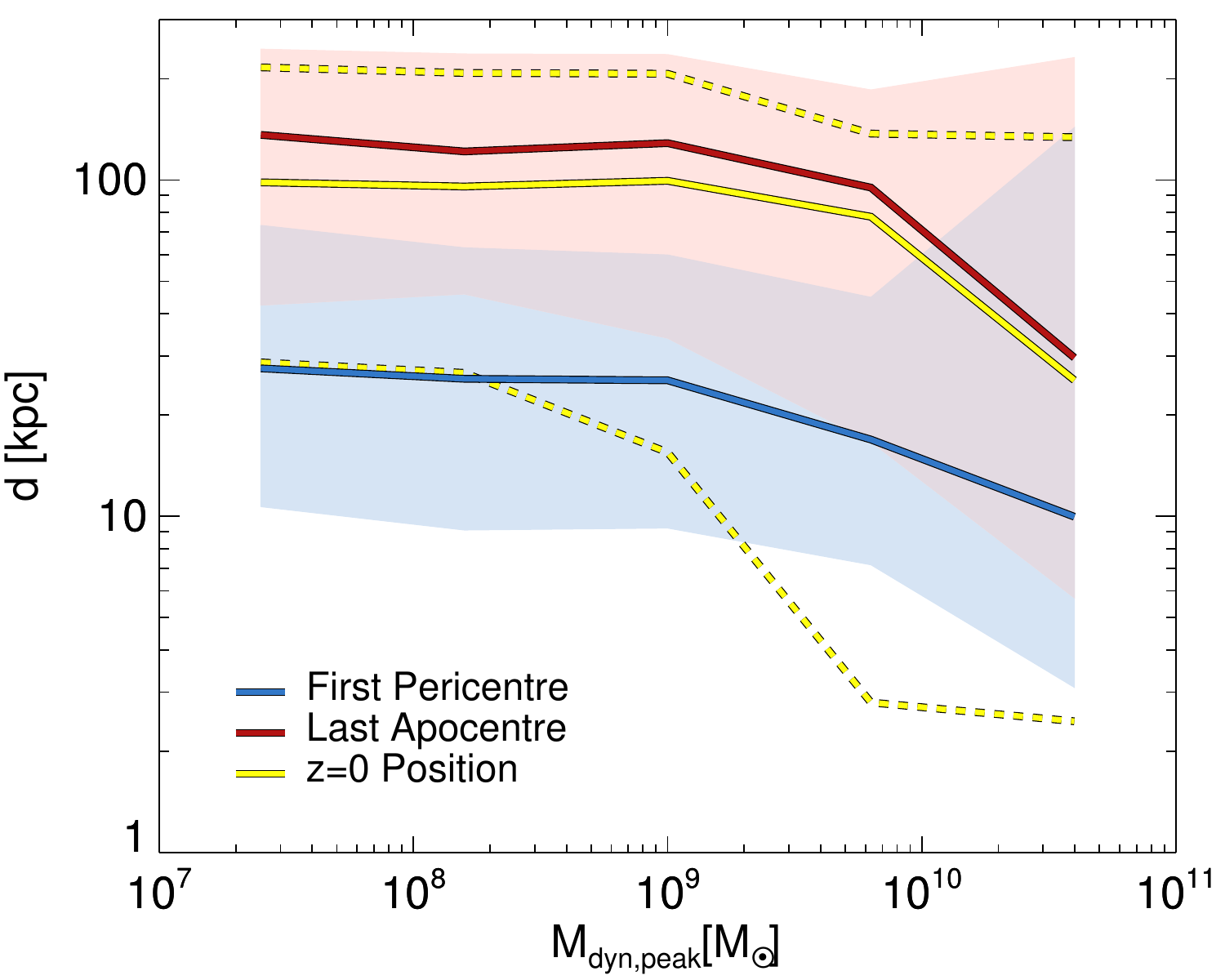}
    \includegraphics[scale=0.35]{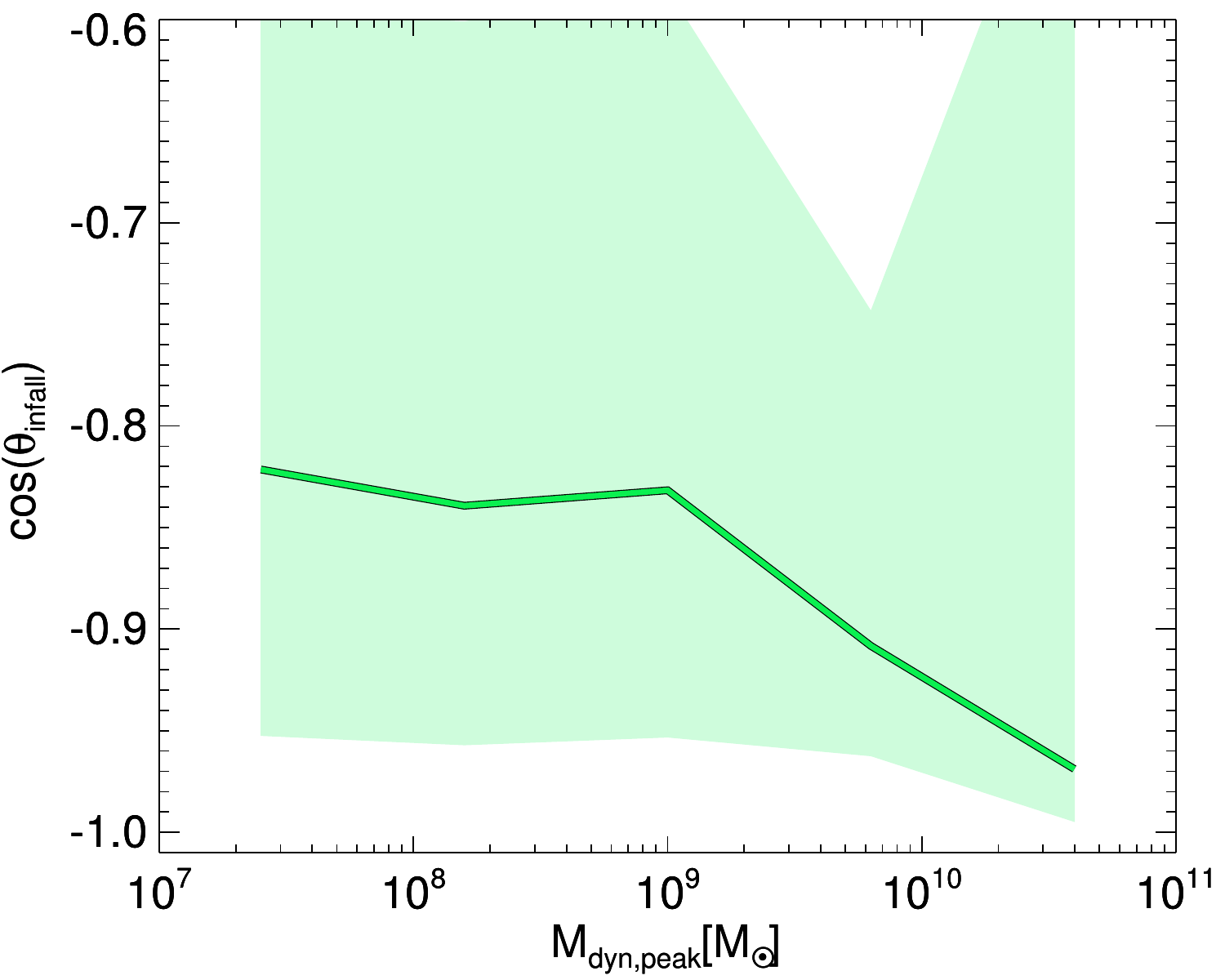}
    \caption{The impact of halo mass on orbital parameters. Left-hand panel: the median first pericentre, last apocentre, and present-day distance-to-host for subhaloes as a function of peak mass shown as blue, red, and yellow solid lines respectively. The 68~per~cent scatter regions for the first pericentre and the last apocentre are shown as shaded regions, while the 68~per~cent scatter in the present day distance is instead indicated with a pair of dashed lines. Right-hand panel: the median infall angle as a function of peak mass, plus the 68~per~cent scatter.}
    \label{fig:dist}
\end{figure*}

\section{Disruption model}
\label{sec:mod}

In this section we describe our model for intuiting the impact of disc disruption on the subhaloes. A comprehensive model for this process would involve growing a disc across cosmic time, accounting for the disc's flattened shape and the adiabatic contraction of the halo \citep{Blumenthal86,Gnedin04}, and negating the risks of spurious disruption due to low mass resolution of the satellites, all of which are beyond the scope of our study. We will therefore return to their likely impact in our conclusions.

Physical disc disruption is a phenomenon that impacts subhaloes whose impact parameter on any orbit is too close to the disc. We therefore consider three types of halo-satellite distance to compare to a notional impact parameter: the present day ($z=0$) position, the most recent apocentre (or `last apocentre'), and the first pericentre after accretion. For the present day position we use the accretion-time most-bound-particle position at $z=0$, as discussed above; we have verified that this is the same as the {\sc subfind}-measured position as and when {\sc subfind} detects the halo at $z=0$. For the last apocentre we take the position at the last snapshot for which the subhalo velocity vector switches from pointing away from the host to pointing towards it. The first pericentre is the position after accretion at which the subhalo velocity angle is first pointing away from the host centre: we  also apply the pericentre-computation method of \citet{Richings20} to further finess this result, although our high time resolution of 13.5~Myr between snapshots means this correction is small.    

One of the key elements of our results will be how these three distance measures change with mass, given that mass is one key property that correlates with how likely a subhalo is to host a luminous galaxy. We therefore compute the median and scatter of present-day position, last apocentre, and first pericentre with the peak dynamical mass of each subhalo, $M_\rmn{dyn,peak}$, which is the peak gravitationally bound mass as determined by the subhalo finder across time. We present these results in the left-hand panel of Fig.~\ref{fig:dist}; in the right-hand panel of the same plot we include the median-plus-scatter of the accretion angle cosine as a function of peak mass to indicate how radial the subhalo orbits are at infall.   


For all three distance measures there is a significant trend towards lower values with higher masses. The average first pericentre for infall dynamical masses of $10^{8}$~$\msun$ is 24~kpc, compared to 12~kpc at $7\times10^{9}$~$\msun$. This result is supported by the infall angle cosine data as a function of $M_\rmn{dyn,peak}$, which show a strong trend for more massive subhaloes to be accreted onto more radial orbits that will therefore bring them closer to the disc. This trend persists with time, as more massive subhaloes also exhibit smaller final apocentres, some fraction of which are within the expected MW disc radius of $\sim20$~kpc \citep[e.g.][]{Cautun20,Lian24,Ou24} for masses $>5\times10^{9}$~$\msun$. The strongest distinction is for the $z=0$ position, although this will include many subhaloes that have in fact merged with the host. Note that for peak masses lower than $6\times10^{9}$~$\msun$ most satellites are located at a distance greater than 80~kpc whereas the opposite is is true for the most massive subhaloes. There are therefore many subhaloes at small masses and large distances, which are difficult to detect if luminous: according to the DES selection function of \citet{DrlicaWagner20}, a galaxy of magnitude $M_{V} = -2.5$ at size 30~pc has an 80~per~cent chance of detection at a distance of $\sim50$~kpc in that survey, compared to 40~per~cent around 100~kpc and is not detectable beyond 128~kpc. Additionally, if subhaloes are rendered dark by reionization feedback \citep{Bullock_00,Benson_02} the true number of satellites would be much smaller. 

For our model of disc disruption we will proceed as follows. We define a parameter, $R_\rmn{DD}$, and any subhalo for which at least one of the first pericentre, last apocentre, or $z=0$ distance is $<R_\rmn{DD}$ is considered to have undergone disruption: here we take disruption to encompass disruption by the tidal field of the notionally adiabatically contracted dark matter halo or by the tidal field of the disc itself, and it will also cover haloes merged by dynamical friction. In the remainder of this paper we will consider five values of $R_\rmn{DD}$: 2, 5, 10, 20, and 30~kpc, and indicate how the change in disc disruption will alter the expected satellite counts and spatial distributions. Note that this model is a semi-analytic formalism applied to $N$-body simulations: in a full hydrodynamical simulation we would obtain an effective $R_\rmn{DD}$ that is a convolution of the disc size, disc mass, and subhalo resolution.   

\section{The 50~kpc survey}
\label{sec:50kpc}

\subsection{Premise}

Whether or not a satellite galaxy is detected in a given survey is based on the convolution of its physical parameters -- stellar mass, surface density, distance from Earth, location on the sky -- and the constraints of the survey: sky coverage, sensitivity of the survey to detect galaxies given the increased difficulty in observing objects at greater distances and / or lower surface densities. In this section we will demonstrate how the interaction of the subhalo mass--orbit distribution on the one hand, and the disc disruption $R_\rmn{DD}$ on the other, will change our inference of the satellite counts in the context of the second survey constraint, that is distance and stellar mass. 

We will consider a fictional survey that is somehow complete for satellites brighter than $M_{V}=-1$ out to a distance of 50~kpc from the centre of the halo and does not consider galaxies of any brightness that are beyond that 50~kpc cutoff. We are therefore using the fictional nature of this survey to neglect effects such as the zone-of-avoidance that hides satellites behind the MW disc and the 8~kpc offset of the Sun from the MW centre. We determine that this survey identifies 20 satellites galaxies within this 50~kpc sphere; for comparison, our current satellite count for the MW in this region is 13, so for demonstration purposes we are assuming there are another 7 undiscovered satellites in this region. 

Our model of the MW system -- including the subhalo spatial distribution, luminous fraction, and disc disruption model -- therefore needs to produce 20 satellites within 50~kpc. If this MW system model produces fewer than 20 subhaloes within this radius, it is ruled out immediately. If instead it produces too many, we will amend the model by introducing a galaxy formation subhalo peak mass threshold to change the luminous fraction. We identify the 20 most massive satellites by $M_\rmn{dyn,peak}$, consider these subhaloes to be luminous and the remainder to be dark due to reionization feedback. This 20th-most-massive subhalo will then set the mass threshold, $M_\rmn{tr}$, for the formation of galaxies throughout the system: $n_\rmn{est}$ is calcuated as the number within 300~kpc for which $M_\rmn{dyn,peak}>M_\rmn{tr}$ and that avoid disc disruption as per the $R_\rmn{DD}$ criterion.

\subsection{Subhalo counts within and outside 50~kpc}

\begin{figure}
    \centering
    \includegraphics[scale=0.34]{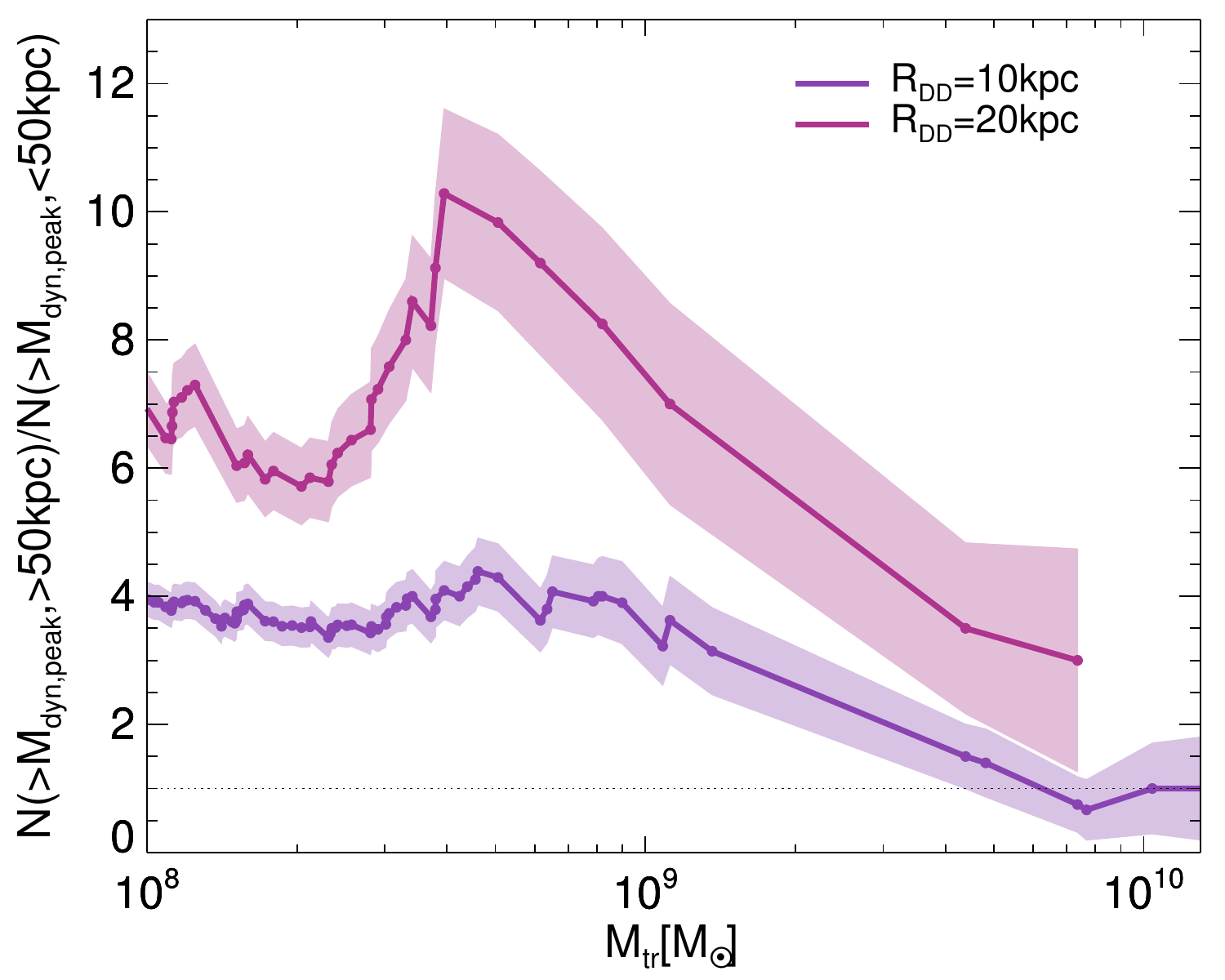}
    \caption{The ratio of the subhalo counts computed for subhalo populations outside and inside 50~kpc when using $R_\rmn{DD}=10$~kpc (purple) and $R_\rmn{DD}=20$~kpc (magenta) as a function of $M_\rmn{tr}$. The shaded regions indicate the $1\sigma$ Poisson uncertainties. }
    \label{fig:innerfrac}
\end{figure}

We will now show how this threshold principle interacts with the spatial distributions of subhaloes and the disc disruption criterion. First, we apply a given value of $R_\rmn{DD}$ to our subhalo population. For those subhaloes that avoid disruption according to this criterion, we order the subhaloes located within 50~kpc at $z=0$ by  $M_\rmn{dyn,peak}$. For each of these inner-halo subhaloes we identify the number of subhaloes more massive than this both outside $50$~kpc and inside $50$~kpc, then compute the ratio of these two counts; this is the ratio of the cumulative $M_\rmn{dyn,peak}$ function at $>50$~kpc to the cumulative $M_\rmn{dyn,peak}$ function at $<50$~kpc. In practice, it states: if my mass threshold $M_\rmn{tr}$ is some value, how many `undetected' luminous satellites may be beyond 50~kpc compared to the `detected' satellitesf within 50~kpc as per our fictional survey? We perform this procedure for both $R_\rmn{DD}=10$~kpc and $R_\rmn{DD}=20$~kpc, then present the results in Fig.~\ref{fig:innerfrac}.   

For all of the $M_\rmn{tr}$ values considered, the ratio of outside-to-inside mass functions is always higher for $R_\rmn{DD}=20$~kpc than for 10~kpc, which is a straightforward statement about how more inner-orbit objects are destroyed when disc disruption extends to larger radii. The two ratios are closest together at very large $M_\rmn{dyn,peak}$, where most subhaloes are on orbits that are primarily within 50~kpc and so the contribution outside is small. The ratio then increases towards lower $M_\rmn{dyn,peak}$ as the intrinsic relative ratio of inner-satellites to outer-satellites increases, but for the stronger disruption model the additional disruption of inner satellites causes a much more dramatic rise: at $7\times10^{8}$~$\msun$ there are four times more objects beyond the survey 50~kpc limit for $R_\rmn{DD}=10$~kpc but nine times more for $R_\rmn{DD}=20$~kpc. At still smaller masses the proportion of satellites orbits changes such that fewer are at risk of disruption and so the fraction drops. Nevertheless, at the rough atomic cooling limit of $10^{8}$~$\msun$ there are still 7 times more objects outside the survey than inside when $R_\rmn{DD}=20$~kpc compared to only 4 times more for $R_\rmn{DD}=10$~kpc. In conclusion, the correlation of orbital parameters with peak mass can lead to a strong sensitivity to disc disruption in inferred satellite counts.

\subsection{Mass functions}

\begin{figure*}
    \centering
    \includegraphics[scale=0.435]{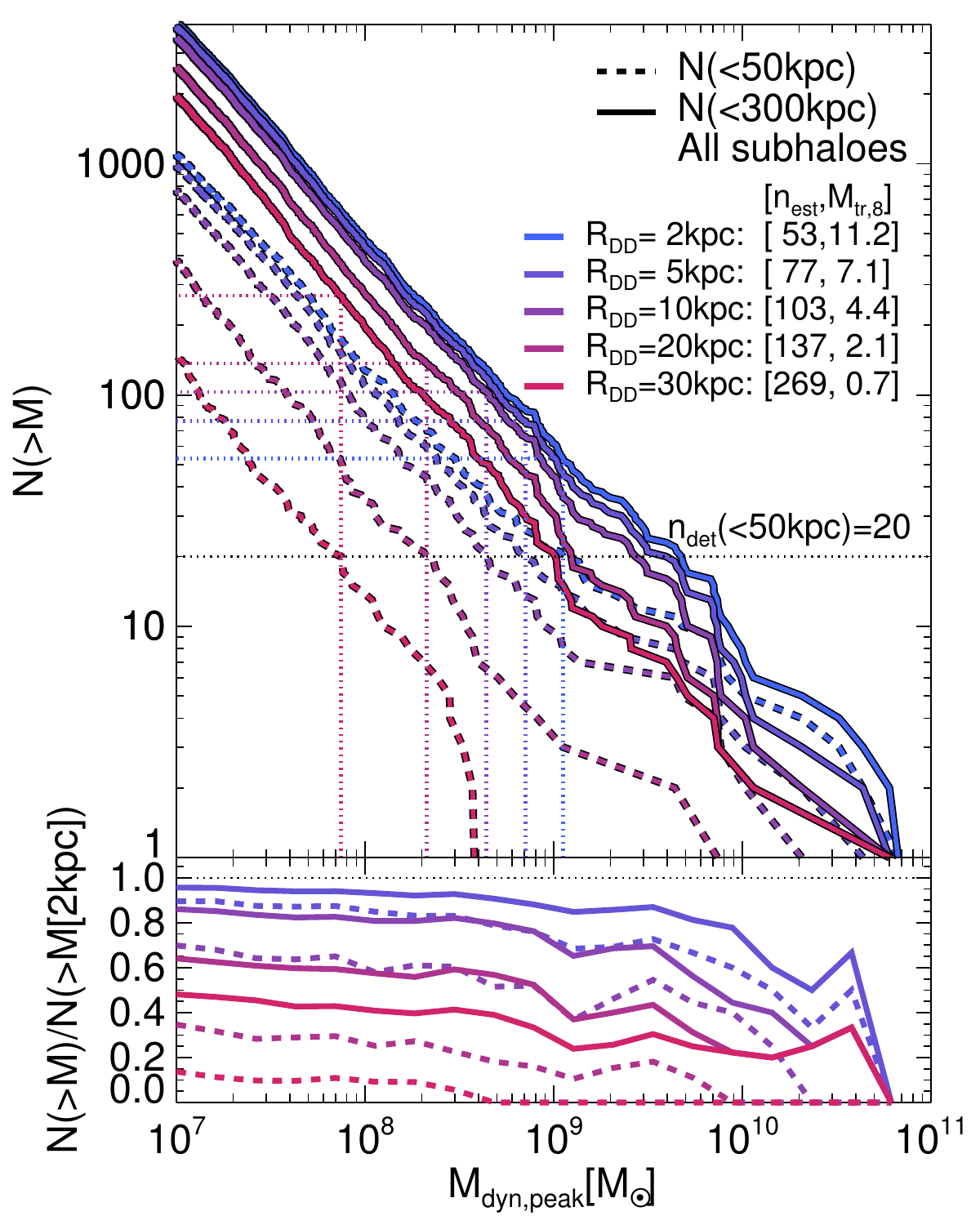}
    \includegraphics[scale=0.435]{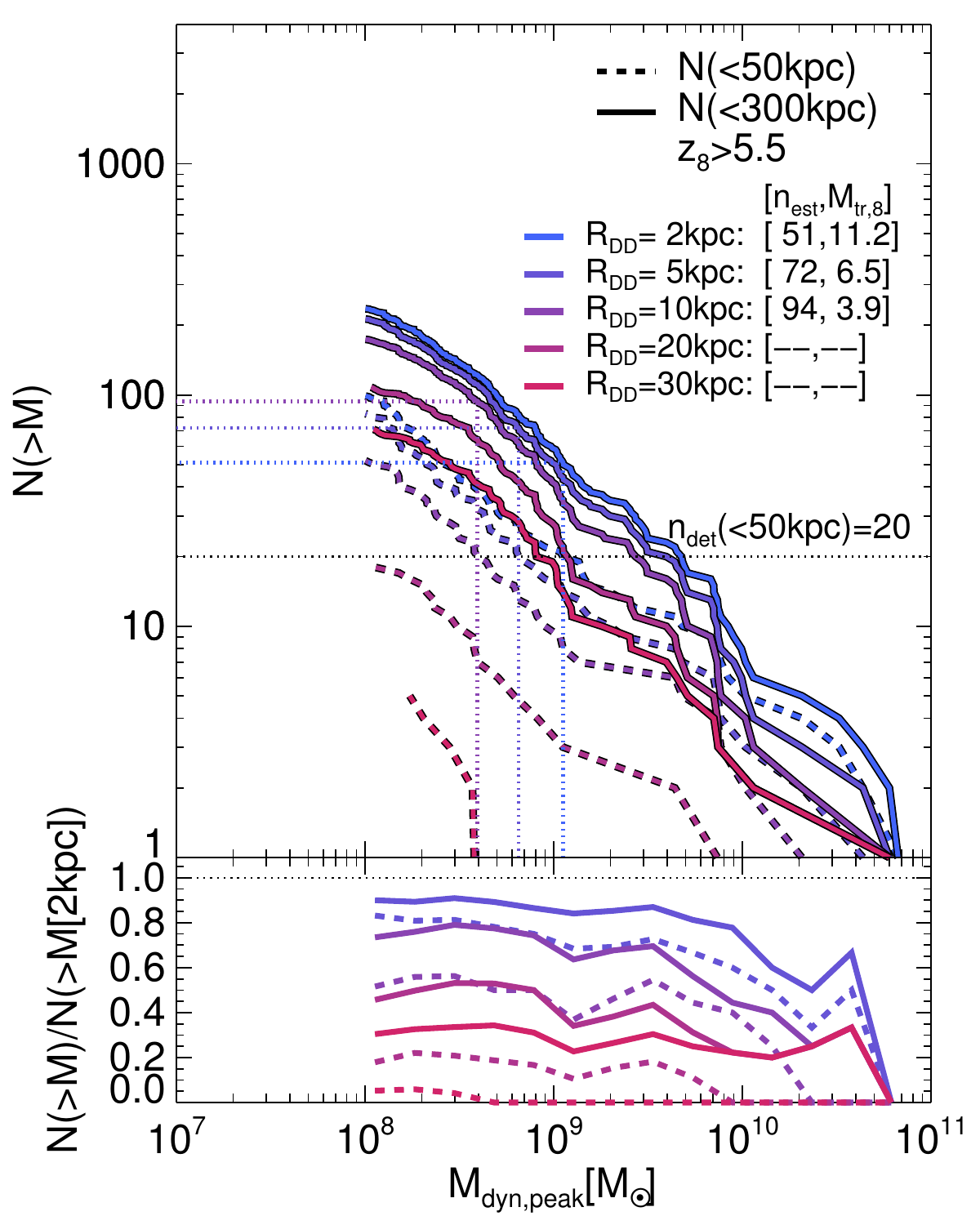}
    \caption{Peak mass functions computed for subhaloes within $50$~kpc (dashed lines) and $300$~kpc (solid lines) assuming different values of $R_\rmn{DD}$. Bluer lines show lower values of $R_\rmn{DD}$ and red lines higher values as indicated in the figure legend. Left-hand panel: all subhaloes are included. Right-hand panel: only subhaloes that have a collapse redshift $z_{8}>5.5$. In both panels the notional survey 50~kpc satellite count of 20 is shown with a horizontal black dotted line. The vertical coloured dotted lines then indicate the threshold mass $M_\rmn{tr}$ for each $R_\rmn{DD}$ value while the corresponding horizontal dotted lines indicate the estimated satellite counts, $n_\rmn{est}$. Both quantities are quoted in the figure legend when there are at least 20 subhaloes within 20~kpc, where we give the threshold mass in units of $10^{8}$~$\msun$, i.e. $M_\rmn{tr,8}\equiv M_\rmn{tr}/(10^{8}$~$\msun)$. In the bottom panels we show the ratio of the mass functions with $R_\rmn{DD}\ge 5$~kpc to their $R_\rmn{DD}=2$~kpc counterparts.}
    \label{fig:mf}
\end{figure*}

We now turn to the mass functions -- across the whole $300$~kpc of the system and within our survey region of $50$~kpc -- for the five values of $R_\rmn{DD}$ that we consider, which we plot in Fig.~\ref{fig:mf}. We identify the $M_\rmn{tr}$ required to obtain our 20 satellites within 50~kpc, then compute the total number of subhaloes with $M_\rmn{dyn,peak}>M_\rmn{tr}$ and note this result in the Figure legend. The process of determining $n_\rmn{est}$ can be followed in the plots by tracing the dotted lines: we start on the right-hand $y$-axis, follow the $n_\rmn{det}=20$ line to the left until it connects with the dashed line of choice, following the corresponding vertical dotted line upwards to its connection with the related solid line, and finally trace the 
companion horizontal dotted line to the left, the value of which is the expected number of luminous satellites within 300~kpc.

One additional consideration for the subhalo mass function is which haloes are are able to cool gas enough to form stars. If one assumes that for dwarf galaxies the predominant cooling mode is through atomic hydrogen, the minimum halo mass to form a galaxy will be in the region of [$10^{8}$,$10^{9}$]~$\msun$ with a dependence on redshift \citep{Bullock_00,Benson_02,BenitezLlambay20}. If instead it is possible to perform cooling through molecular hydrogen cooling \citep[see e.g.][and references therein]{Nadler25}, the halo mass limit would be $10^{6}$~$\msun$, and well below the mass resolution limit of our simulation. We therefore produce two sets of mass functions for Figure~\ref{fig:mf}: one in which we include all subhaloes, and a second in which we exclude subhaloes that are not likely to meet the atomic cooling standard. A careful treatment of atomic cooling would involve a mass threshold that evolves with redshift \citep{BenitezLlambay20} or even applies ray tracing of radiation packets \citep[e.g.][]{Shen24}; in the spirit of our toy model approach, we instead simply remove all of the haloes that fail to achieve the approximate atomic cooling mass of $10^{8}$~$\msun$ before the end of reionization, which we take to be $z=5.5$ \citep[see e.g.][]{Bosman22}. 

We will first consider the results that are not restricted to atomic cooling. Increasing $R_\rmn{DD}$ from 2~kpc to 5~kpc decreases the total number of subhaloes within 300~kpc by only 15~per~cent for $M_\rmn{dyn,peak}<10^{9}$~$\msun$, as indicated in the bottom panel. However, the number of satellites within the inner 50~kpc drops by 25~per~cent in this range. $M_\rmn{tr}$ falls from $\sim1\times10^{9}$~$\msun$ to $7\times10^{8}$ ~$\msun$, and by our $M_\rmn{tr}$ luminous fraction model the number of satellites increases from 53 to 77. Therefore, even as the total number of {\it subhaloes} within 300~kpc has decreased by 15~per~cent, the number of anticipated {\it luminous satellites}, i.e. $n_\rmn{est}$, has increased by 45~per~cent. This pattern continues to larger $R_\rmn{DD}$, with stronger suppression in the inner region compared to the halo as a whole leading to higher and higher $n_\rmn{est}$, from $n_\rmn{est}=103$ at $R_\rmn{DD}=10$~kpc to  $n_\rmn{est}=137$ at $R_\rmn{DD}=20$~kpc and 269 at $R_\rmn{DD}=30$~kpc. It is also the case that the suppression in central  subhalo counts gets stronger with $M_\rmn{dyn,peak}$ as shown above, forcing $M_\rmn{tr}$ to ever lower values where the total number of haloes across the host is much higher.

The introduction of atomic cooling by construction forces $M_\rmn{tr}$ to marginally smaller values, including from $4.4\times10^{8}$~$\msun$ to $3.9\times10^{8}$~$\msun$ at $R_\rmn{DD}=10$~kpc, or a drop of 11~per~cent. The $R_\rmn{DD}=20$~kpc and $R_\rmn{DD}=30$~kpc models fail to generate enough haloes to match the 50~kpc region 20 satellite target. In these models, it is therefore necessary to consider molecular cooling as argued elsewhere \citep[e.g.][]{Graus:2019}. For $R_\rmn{DD}\leq10$~kpc $M_\rmn{tr}$ is not strongly affected, yet as $R_\rmn{DD}$ increases from 2~kpc through 5~kpc to 10~kpc the total number of subhaloes with $M>M_\rmn{tr}$ increases more slowly than when no atomic cooling cut is applied despite the decrease in $M_\rmn{tr}$. This is because halo collapse time correlates with distance to the central galaxy \citep{Lovell24} and so at fixed mass the fraction of subhaloes outside 50~kpc that pass the atomic cooling redshift cut is lower than those within 50~kpc.

\subsection{Radial distributions}

\begin{figure}
    \centering
    \includegraphics[scale=0.345]{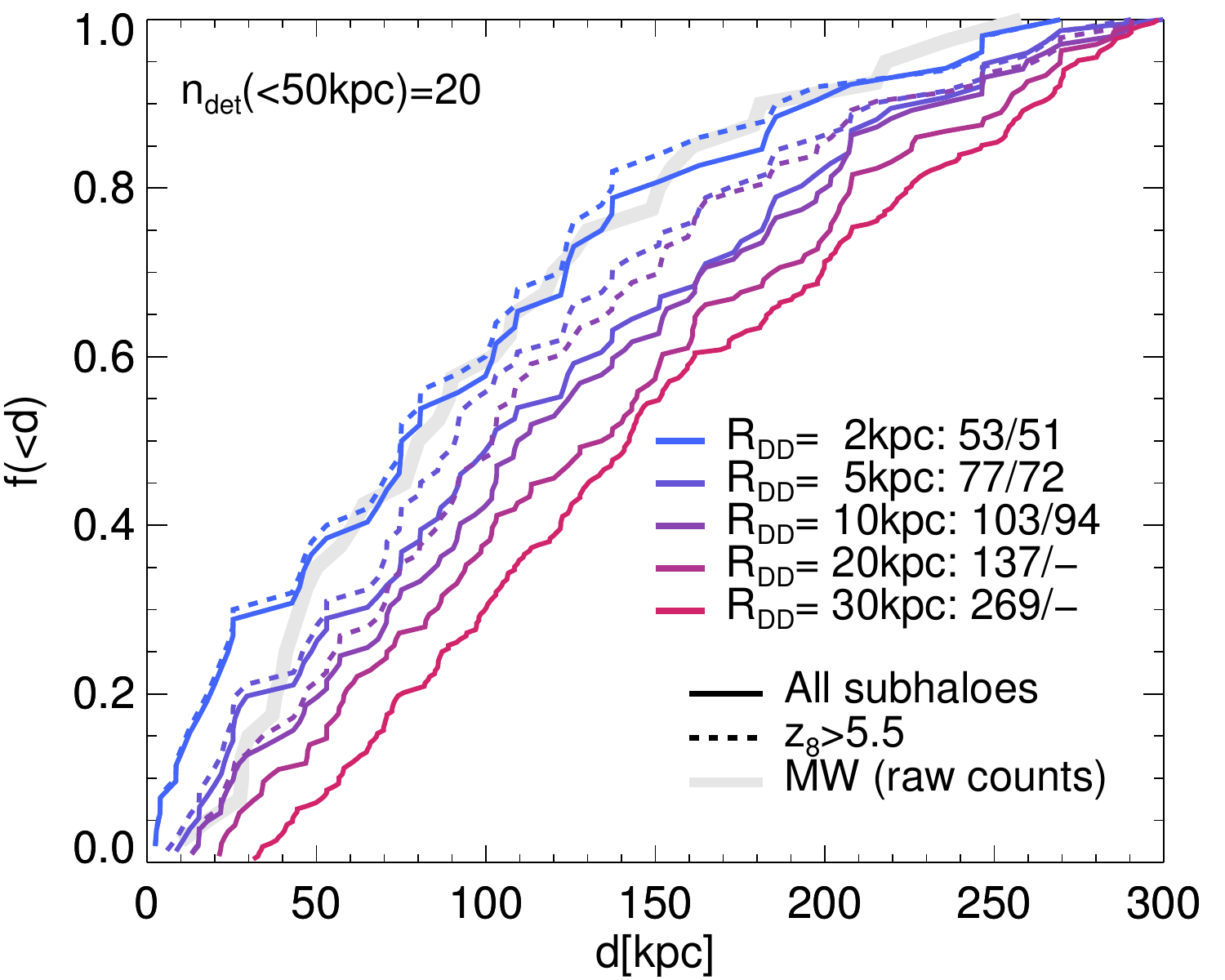}
    \caption{Radial distributions for the survey-selected subhaloes while varying $R_\rmn{DD}$ and the adoption or otherwise of the $z_{8}>5.5$ cut. $R_\rmn{DD}$ is indicated by colour as shown in the legend. The application of $z_{8}>5.5$ as dashed lines compared to solid where it is not applied; $n_\rmn{est}$ in these models is indicated to the right and left of the slash in the legend respectively. The grey line indicates the spatial distribution of the observed MW satellites.}
    \label{fig:rad}
\end{figure}

The final piece of the puzzle that we consider in this section is the radial distribution of satellites. For each of our models that generate enough $<50$~kpc subhaloes to match the 20 survey satellites -- that is, all except for $R_\rmn{DD}\ge20$~kpc when our atomic cooling criterion is enforced -- we select all of the subhaloes that have $M_\rmn{dyn,peak}>M_\rmn{tr}$ and plot their normalised radial distributions, $f(<d)$ in Fig.~\ref{fig:rad}. We also include the current raw counts for the known MW satellites: we do not attempt to correct for any of the observational effects here. 

There is a clear trend for smaller $R_\rmn{DD}$ values to lead to more concentrated distributions: at 50~kpc the two $R_\rmn{DD}=10$~kpc models predict a fraction of $f\sim0.2$ whereas for $R_\rmn{DD}=30$~kpc $f\sim0.1$. All of the models with $R_\rmn{DD}>5$~kpc sit below the current observations at distances less than $150$~kpc, although it is a reasonable assumption that future discoveries of as yet undetected, faint satellites will be made further out in the halo, therefore the observational line may come down to meet the model predictions. The $R_\rmn{DD}\leq5$~kpc models are significantly more concentrated than the observations, which indicates that potentially these models underestimate disc disruption. The impact of the atomic cooling is to make the three viable models more concentrated, as the condition removes outlying, late forming small subhaloes from the galaxy formation pool. 

\begin{figure*}
    \centering
    \includegraphics[scale=0.6]{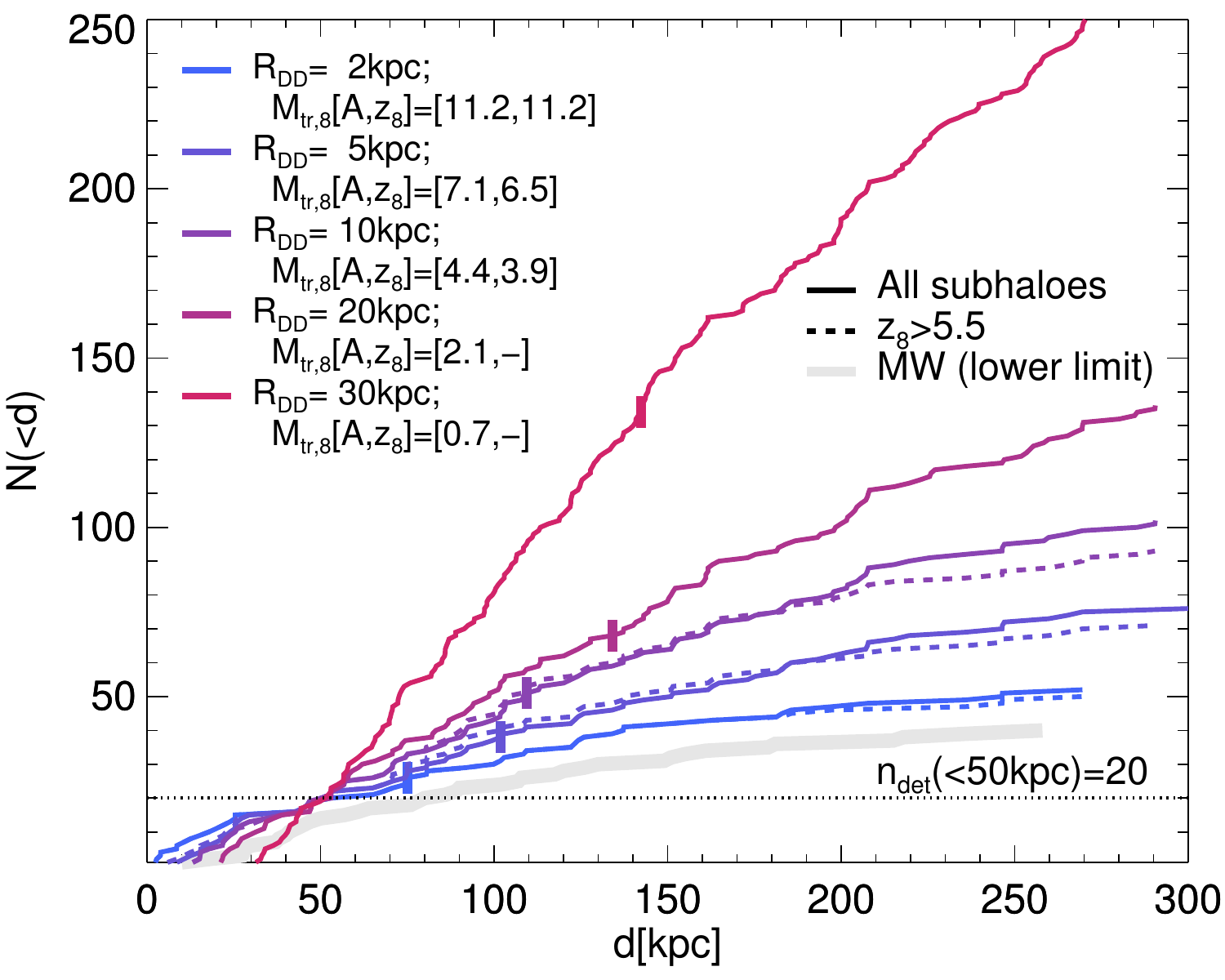}
    \caption{Radial distributions for the survey-selected subhaloes while varying $R_\rmn{DD}$ and the adoption or otherwise of the $z_{8}>5.5$ cut. $R_\rmn{DD}$ is indicated by colour as shown in the legend. The application of $z_{8}>5.5$ is shown as dashed lines compared to solid where it is not applied. In the latter case we denote the radius that includes half of the satellites as a vertical bar. The mass threshold $M_\rmn{tr}$ is indicated in the figure legend in units of $10^{8}$~$\msun$, i.e. $M_\rmn{tr,8}\equiv M_\rmn{tr}/(10^8$~$\msun)$, first when using all subhaloes ([A]) and second for the atomic cooling model ([$z_{8}$]). The grey line indicates the spatial distribution of the observed MW satellites. Note that all models pass through the (50,20) point by construction.}
    \label{fig:radABS}
\end{figure*}

We summarize this section with the absolute number counts as a function of radius in Fig.~\ref{fig:radABS}. This is the same dataset as shown in Fig,~\ref{fig:rad} but without dividing through by $n_\rmn{est}$, which therefore enables us to show how many additional satellites one would expect to see with distance. It shows that the requirement to find 20 satellites within 50~kpc forces the radial distribution to pivot about this point. The destruction of high-mass subhaloes within 50~kpc requires that more lower mass subhaloes be luminous, therefore the loss of high-mass subhaloes to disc disruption is more than balanced by putting satellites into many more subhaloes at larger radii as $R_\rmn{DD}$ increases. The radial concentration drops simultaneously, with $f=0.5$ occurring at $103$~kpc for $R_\rmn{DD}=5$~kpc but $143$~kpc for $R_\rmn{DD}=30$~kpc The number is primarily impacted by disc disruption up to 150~kpc, beyond which atomic cooling effects become important. At smaller radii the atomic cooling models have slightly higher satellite counts than their counterpart models due to the lower $M_\rmn{tr}$, but beyond 150~kpc the haloes are younger and therefore not present in the atomic-cooling sample, so the total count is still lower than when cooling considerations are ignored. 

We therefore have a straightforward, qualitative picture for how the disc disruption impacts the number and distribution of satellite galaxies. As the disruption becomes stronger, that is, satellites that are progressively further from the host centre undergo disruption, the model is forced to populate lower mass haloes in order to match the `observed' number of nearby satellites; the mass threshold $M_\rmn{tr}$ drops from $\sim10^{9}$~$\msun$ for $R_\rmn{DD}=2$~kpc to $\sim4\times10^{8}$~$\msun$ at $R_\rmn{DD}=10$~kpc, and continuing to below the atomic cooling threshold at still larger stripping radii. Progressively lower mass haloes are more numerous at large radii beyond the detection threshold of current surveys, so we anticipate that satellite populations are both larger and less radially concentrated when disc disruption is stronger. The number of expected satellites increases from 53 at $R_\rmn{DD}=2$~kpc to 103 at $R_\rmn{DD}=10$~kpc, with a suppression of up to another 10 satellites by atomic cooling limits; models that rely on molecular cooling to match the inner halo counts may well require $>200$ satellites in total.

\section{DES/Pan-STARRS}
\label{sec:despan}

In Section~\ref{sec:50kpc}, we effectively adopted a simple survey selection function (100~per~cent completeness within 50~kpc) and galaxy formation model (place an infinitely bright galaxy in each halo with peak mass greater than $M_\rmn{tr}$) to examine the interplay between disc disruption ($R_\rmn{DD}$) and the inferred galaxy-halo connection.
We now relax some of these simplifications for a more realistic setup based on the observed MW satellites, heavily inspired by \citet{Nadler20}\footnote{Our calculations in this section utilize \href{https://github.com/eonadler/subhalo_satellite_connection/tree/master/1912.03303}{the code release} corresponding to \citet{Nadler20}, except where otherwise noted in the main text.}.

Using the same simulated halo introduced in Section~\ref{sec:sim}, we populate subhaloes with luminous satellites using a modified version of the galaxy formation model detailed in Section 4 of \citet{Nadler20}.
They list the parameters, along with constraints on those parameters, in their Table~1; we adopt the same parameters for the rest of this section.
The following aspects of the model remain the same:
\begin{itemize}
    \item Luminosities: haloes of a given peak circular velocity, $V_\text{peak}$, are assigned an absolute V-band magnitude, $M_V$, according to a relation constrained by the GAMA survey \citep{Loveday:2015, Geha:2017} for systems brighter than $M_V = -13$ and extrapolated to fainter systems. The extrapolation assumes a faint-end slope of the luminosity function, $\alpha$, and a lognormal scatter, $\sigma_M$, in luminosity at fixed $V_\text{peak}$.
    \item Galaxy formation efficiency: haloes are assigned a probability of forming a galaxy that varies smoothly with peak halo mass. This probability is modeled as a Gaussian error function with a 50\% occupation rate at $\mathcal{M}_{50}$ and a width of $\sigma_\text{gal}$ (see Equation 3 of \citealp{Nadler20}).
\end{itemize}

We apply the following modifications to the model:
\begin{itemize}
    \item Galaxy size: we assign mean predicted 3D stellar sizes at accretion using the $r_\text{1/2} = 0.015\ R_\text{200c}$ relation from \citet{Kravtsov:2013}. As in \citet{Nadler20}, sizes are drawn from a lognormal distribution with $\sigma_{\log R}$, equated to azimuthally averaged projected half-light radii, and assumed not to change between accretion and the present day.
    \item Baryonic disruption: we adopt the model described in Section~\ref{sec:mod}. Satellites with $R_\rmn{DD}$ below a certain value are assigned $p_\text{disrupt} \equiv 1$, otherwise $p_\text{disrupt} \equiv 0$.
    \item Numerical disruption: as discussed in Section~\ref{sec:sim}, we place all satellites at the present-day location of the most-bound particle at accretion, essentially ignoring if a subhalo is fully disrupted or survives to the present day (according to the halo catalogues). Their disruption probability is governed by the same $R_\rmn{DD}$ formalism as the previous bullet point.
\end{itemize}
All of these model prescriptions are either probabilistic or stochastic in nature, such that for a fixed set of model parameters there will be scatter in the predicted satellites of a given subhalo population.

After generating the total population of satellites using this formalism, we then determine how many of those satellites are detected in a combined mock DES and PS1 survey.
We choose random observer locations 8~kpc away from the halo centre, then rotate coordinates so that the most massive subhalo is located at the on-sky position of the LMC.
We then employ the DES and PS1 survey selection functions derived in \citet{DrlicaWagner20}, which predict the satellite detection probability given absolute magnitude, heliocentric distance, half-light radius, and sky position.
The number of detectable satellites is:
\begin{equation}
    N_\text{sat,det} \equiv \sum_i p_\text{occ,i} \times (1-p_\text{disrupt,i}) \times p_\text{det,i},
\end{equation}
where for the $i$-th subhalo, $p_\text{occ}$ is the mass-dependent probability that the subhalo is occupied by a galaxy, $p_\text{disrupt}$ is the probability the satellite is disrupted due to the host disc, and $p_\text{det}$ is the probability that the satellite is detected in either DES or PS1.
If a satellite falls in the overlap between the two surveys, we use the DES detection probability.

Since our aim is to add a few layers of realism to the toy setup in Section~\ref{sec:50kpc}, rather than perform a full fit of this model to the MW satellite data, we fix all galaxy formation model parameters to values consistent with the constraints reported in \citet{Nadler20}.
Echoing the exercise in Section~\ref{sec:50kpc}, we then adopt one value for $R_\rmn{DD}$ and vary the galaxy occupation fraction parameter $\mathcal{M}_{50}$, which is the peak halo mass at which 50~per~cent of haloes host a galaxy.
We lower $\mathcal{M}_{50}$ until the number of detected satellites $N_\text{sat,det}$ is approximately 34, the number of confirmed MW satellites in both surveys using the census in \citet{DrlicaWagner20} adopted by \citet{Nadler20}.

\begin{figure*}
    \centering
    \includegraphics[width=1.0\textwidth]{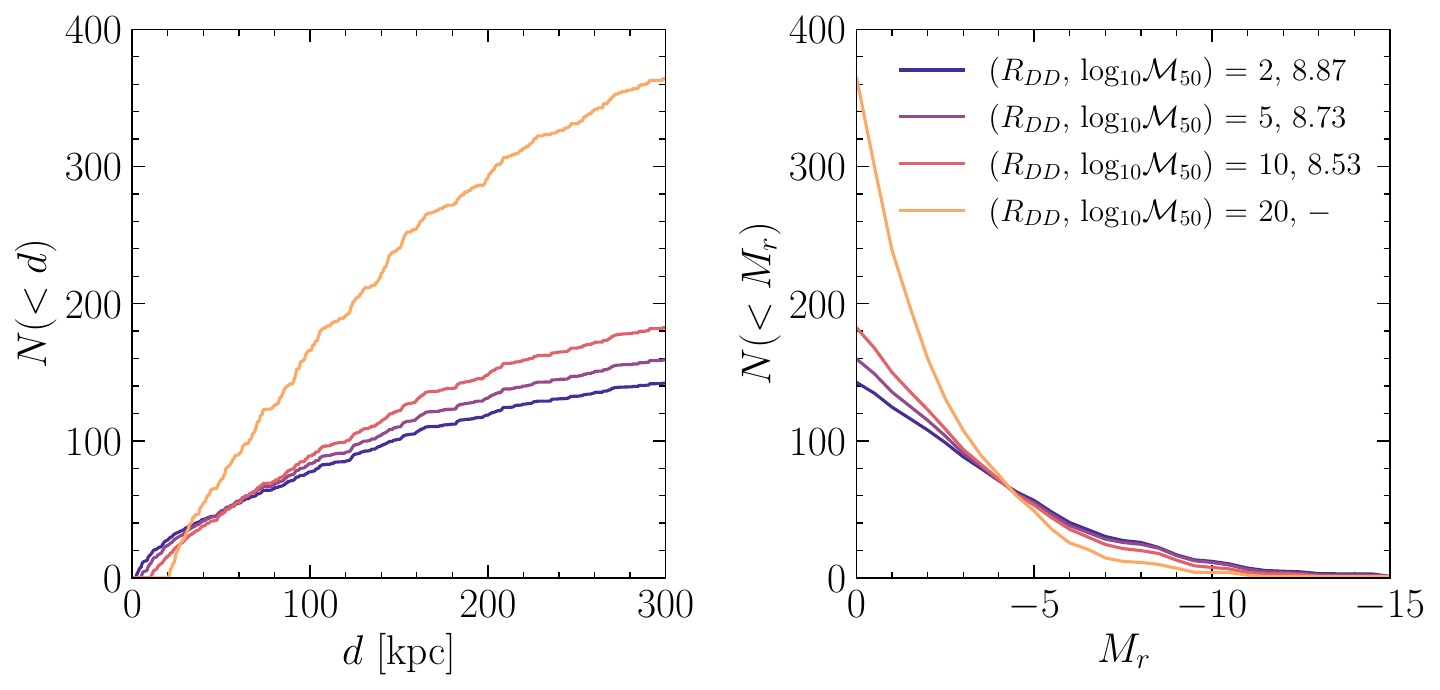}
    \caption{Radial distributions (left panel) and luminosity functions (right panel) using the DES+PS1 selection functions \citep{DrlicaWagner20} and a galaxy formation model similar to that in \citet{Nadler20}.
    Each $R_\rmn{DD}$ has a corresponding galaxy formation threshold mass $\mathcal{M}_{50}$ indicated in the figure legend, in units of kpc and $\msun$ respectively.
    Note that all models shown provide approximately 34 detectable satellites within 300~kpc.}
    \label{fig:nadlermodel}
\end{figure*}

The top-level results of this exercise are presented in Figure~\ref{fig:nadlermodel}, where we show the radial distributions (left panel, analogous to Figure~\ref{fig:radABS}) and luminosity functions (right panel, partially analogous to Figure~\ref{fig:mf}) for the total satellite population with a given $R_\rmn{DD}$.
In agreement with Section~\ref{sec:50kpc}, we find that enhanced disruption rates (increasing $R_\rmn{DD}$) requires populating lower-mass subhaloes (decreasing $\mathcal{M}_{50}$) in order to provide $N_\text{sat,det} \sim 34$ detectable satellites.
This results in more satellites overall, with a less centrally concentrated radial distribution about the host.

We initially adopted the fiducial model parameters reported in Figure~5 of \citet{Nadler20}, but found it difficult to provide enough detectable satellites for larger values of $R_\rmn{DD}$.
Instead, we adopted parameters within the 95~per~cent~constraints reported in their Table~1, always biasing towards producing more and brighter satellites.
The fixed parameters are the following: $\alpha = -1.46$, $\sigma_M = 0.05$\footnote{This scatter in the $M_\star-V_\text{peak}$ relation is much smaller than the values considered in \citet{Santos-Santos:2022} motivated by cosmological hydrodynamical simulations, mitigating the increase they see in the satellite luminosity function in models with such scatter (their Figure~11).}, $\sigma_\text{gal} = 0.67$, $\sigma_{\log R} = 0.33$.
This configuration provided enough detectable satellites to provide an $\mathcal{M}_{50}(R_\rmn{DD})$ constraint for values of $R_\rmn{DD}$ up to $10 \kpc$.
Further increasing $R_\rmn{DD}$ to $20 \kpc$ provided $N_\text{sat,det} = 33.6 \pm 3.0$ satellites if $\mathcal{M}_{50}$ is set well below our $M_\text{dyn,peak}$ resolution limit.
Effectively, this means that every subhalo is populated with a satellite and the limit on the number of detectable satellites arises from a combination of $p_\text{disrupt}$ and $p_\text{det}$.
The largest value of $R_\rmn{DD} = 30 \kpc$ considered in Section~\ref{sec:50kpc} removes too many massive subhaloes to provide enough detectable satellites under this configuration of model parameters.

We caution that this analysis is far from a fit to the observed MW satellite data.
In particular, the most massive subhalo in our simulation is neither as massive or as close as the LMC.
The number of satellites is the DES footprint is likely enhanced due to the LMC bringing in a number of its own satellite galaxies \citep[e.g.][]{Nadler20, Patel:2020, Santos-Santos:2021, Vasiliev:2023}.
This is why we only consider the total number of detected satellites in the combined DES and PS1 footprints, rather than the two footprints individually or their relative balance simultaneously.

Finally, we note that both our exercise and \citet{Nadler20} assume that satellites either have their peak stellar mass or are fully disrupted and removed from the analysis.
This is in stark contrast to recent analyses of disrupting satellites in cosmological simulations \citep{Panithanpaisal:2021, Riley:2025, Shipp:2025, Pathak:2025}, where many surviving satellites have experienced substantial amounts of stellar mass loss due to tidal forces.
While these simulations have so far focused on satellites with peak stellar mass $M_\star \gtrsim 10^5 \msun$, it is unclear if these lessons apply to lower mass ultra-faints which are the dominant contributor to satellite counts analyses.
We encourage further work in this regard.

In summary, adopting a more realistic setup for populating and detecting MW satellites reinforces the qualitative results of this work.
The number of detected MW satellites is sensitive to the magnitude of disruption from the host potential.
Increasing the rate of disruption, while keeping the number of detectable satellites fixed, requires populating less massive subhaloes, resulting in more satellites overall and a less concentrated distribution about the host.
Adopting a single model for disruption, without allowing for some flexibility or modeling uncertainty, may result in empirical galaxy formation models that are overconfident about their underlying model parameters (e.g. stellar mass-halo mass relation, occupation fraction).

\section{Conclusions}
\label{sec:conc}

The MW satellite count remains a crucial test for any viable dark matter model. Estimating this count relies on a combination of observations and simulation priors, where the former includes the sensitivity and size of our surveys and the latter includes the radial distribution of host dark matter subhaloes and the disruption of those subhaloes. We have highlighted from the literature a particular issue that the impact of disc disruption is not clear, and as simulation resolution improves satellite galaxies become more resilient, restricting disruption to galaxies ever closer to the galactic centre \citep{Grand21}.

In this study we develop a model to show how uncertainties in susceptibility for satellite disruption may propagate through to the eventual satellite count prediction. We first perform a zoomed $N$-body simulation of a MW-mass halo at very high time resolution in order to be able to track subhalo orbits. We generate merger trees to compute these orbits from first infall and identify each subhalo's first pericentre, last apocentre, and present day position. We then introduce a disc disruption parameter -- $R_\rmn{DD}$ -- such that any subhalo for which any one of the first pericentre, last apocentre, or present day position was smaller than $R_\rmn{DD}$ is considered disrupted and so removed from the list of subhaloes eligible to contribute to the satellite count.  

In Fig.~\ref{fig:dist} we show how the subhalo orbital parameters vary with mass, with more massive subhaloes accreted onto more radial orbits, leading to smaller pericentres and apocentres especially in combination with dynamical friction. Given that it is the more massive subhaloes that also have the highest probability of forming a luminous satellite, we showed that large $R_\rmn{DD}$ would likely require star formation in subhaloes of lower mass, which have a greater relative abundance at large radii; we demonstrate this latter point explicitly in Fig.~\ref{fig:innerfrac}.

We then conceive of a fictional survey that is complete in all directions out to a radius of 50~kpc, in order to illustrate how changing $R_\rmn{DD}$ alters the expectations for the total ($<300$~kpc) galaxy counts, $n_\rmn{est}$. In Fig.~\ref{fig:mf} the number of `observed' satellites within $50$~kpc is set to 20. If we assume that the 20 most massive surviving subhaloes within $50$~kpc are luminous then this determines a minimum threshold mass for subhalos to host luminous satellites, $M_\rmn{tr}$. Counting those satellites, as well as those that above this mass threshold in the outer halo while assuming the rest are dark, leads to $n_\rmn{est}=53$ at $R_\rmn{DD}=2$~kpc and to $n_\rmn{est}=269$ at $R_\rmn{DD}=30$~kpc, and the satellite radial distribution concentration significantly decreases (Fig.~\ref{fig:rad} and Fig.~\ref{fig:radABS}). When imposing a $M_\rmn{tr}$ of $10^8$~$\msun$, in agreement with general atomic cooling arguments, $R_\rmn{DD}=30$~kpc generates only half of the required subhaloes within 50~kpc. Finally, we apply a more realistic survey footprint from the combined PanSTARRS and DES surveys in the same process and find similar increases in number counts / decreases in radial concentrations with increasing disc disruption radius.

We summarize the results of our toy model experiments as follows. Observational surveys are best placed to detect satellites that are luminous, compact, and located at small distances from Earth. In the absence of any disruption by the disc, the population of subhaloes within 50~kpc of the host haloes far exceeds the number of detected satellite galaxies, which is the original missing satellite problem \citep{Klypin99,Moore99}. We then require heating from reionisation to prevent galaxy formation in all but the most massive subhaloes in order to match the observed satellite counts. These subhaloes are biased towards the halo centre due to a preference for initial radial orbits and due to dynamical friction. We therefore obtain a population of luminous satellites that is concentrated towards the halo centre. When disc disruption is introduced for the tightest orbits, the most massive subhaloes are preferentially destroyed, and the previous reionisation mass threshold becomes too strong to generate enough satellites to match the counts. It is therefore necessary to reduce that threshold to lower masses in order to match the observed satellites. A side effect of lowering the mass threshold is to `illuminate' a subhalo population that is less biased towards the inner radii, so we infer a larger population of satellites in the outer halo beyond what can be detected in the survey. This inference of a population of faint, distant satellites in low mass subhaloes increases as stronger disc disruption destroys progressively less massive subhaloes in the halo centre and so requires still less massive haloes to be luminous.

We have demonstrated this dependence of satellite estimates on the disc disruption in the straightforward case of an isolated dark matter-only MW-mass halo and disc disruption that occurs in a spherical region. There are therefore many simplifying assumptions in our model, and instead a comprehensive, first-principles model will be required to establish how any uncertainty in the disruption efficiency propagates into uncertainty in the in the complete MW satellite count. The most pressing example is the use of a more accurate MW halo-analogue as discussed above. The presence of an LMC at a distance of 50~kpc from the halo centre may introduce a new population of late-accreted satellites that have not had time to be disrupted. These new satellites in the halo centre would potentially allow the $R_\rmn{DD}=30$~kpc model to host a sufficient number of $<50$~kpc satellites to match the known satellites. The presence of an additional overdensity in the neighbourhood of the M31 galaxy may also make a difference, as would the changing orientation of the MW disc. These efforts are already underway \citep[e.g.][]{Buch24}, our caution is to account for potential uncertainties in disruption rates, both physical -- by the host disc -- and numerical  -- from halo finder errors, and from artificial disruption \citep{SantosSantos25}. Such concerns are essential for modelling MW satellite counts in the upcoming era of Rubin LSST \citep[e.g.][]{Tsiane:2025}.

\section*{Acknowledgements}

We thank Ethan Nadler and Nora Shipp for enlightening discussions.
This research made extensive use of \href{https://arxiv.org/}{arXiv.org} and the Science Explorer, funded by NASA under Cooperative Agreement 80NSSC21M00561.

AHR was supported by a fellowship funded by the Wenner Gren Foundation, a Research Fellowship from the Royal Commission for the Exhibition of 1851, and by STFC through grant ST/T000244/1. ISS acknowledges support from the European Research
Council (ERC) Advanced Investigator grant to C.S. Frenk, DMIDAS
(GA 786910) and from the Science and Technology Facilities
Council [ST/P000541/1] and [ST/X001075/1]. This work used the DiRAC@Durham facility managed by the Institute for Computational Cosmology on behalf of the STFC DiRAC HPC Facility (www.dirac.ac.uk). The equipment was funded by BEIS capital funding via STFC capital grants ST/K00042X/1, ST/P002293/1, ST/R002371/1 and ST/S002502/1, Durham University and STFC operations grant ST/R000832/1. DiRAC is part of the National e-Infrastructure.

\section*{Data Availability}

Anyone interested in using the data used in this paper should contact MRL at m.r.lovell@durham.ac.uk.

\section*{Software}

This work made use of the following software packages: \texttt{astropy} \citep{astropy:2013,astropy:2018,astropy:2022}, \texttt{matplotlib} \citep{Hunter:2007}, \texttt{numpy} \citep{numpy}, \texttt{pandas} \citep{mckinney-proc-scipy-2010,pandas_17229934}, \texttt{python} \citep{python}, \texttt{scipy} \citep{2020SciPy-NMeth,scipy_11255513}, \texttt{Cython} \citep{cython:2011}, \texttt{h5py} \citep{collette_python_hdf5_2014,h5py_7560547}, and \texttt{xgboost} \citep{2016arXiv160302754C}.

This research has made use of the Astrophysics Data System, funded by NASA under Cooperative Agreement 80NSSC21M00561.

Parts of the results in this work make use of the colormaps in the CMasher package \citep{2020JOSS....5.2004V,CMasher_14186007}

Some of the results in this paper have been derived using \texttt{healpy} and the HEALPix package\footnote{http://healpix.sourceforge.net} \citep{Zonca2019, 2005ApJ...622..759G,healpy_17786647}

Software citation information aggregated using \texttt{\href{https://www.tomwagg.com/software-citation-station/}{The Software Citation Station}} \citep{software-citation-station-paper,software-citation-station-zenodo}.



\bibliographystyle{mnras}








\bsp	
\label{lastpage}
\end{document}